\documentclass{JFM-FLM_Au}

\usepackage{bm}

\lefttitle{O. Abukabsha, S. Gsell and M. Brandenbourger}
\righttitle{Journal of Fluid Mechanics}

\title{On the rectification of oscillatory flows by flexible leaflets in a confined geometry}

\author{Omar Abukabsha\aff{1}, Simon Gsell\aff{1} \and Martin Brandenbourger\aff{1}}

\affiliation{\aff{1}Aix Marseille Univ, CNRS, Centrale Med, IRPHE, Marseille, France}

\corresau{Martin Brandenbourger, \email{martin.brandenbourger@univ-amu.fr}}

\begin{document}

\maketitle

\begin{abstract}
Inspired by biological systems, extensive research has explored how fluid-structure interactions in compliant channels and confined geometries control fluid transport. While local nonlinearities can be induced by individual components, arranging these elements into larger architectures gives rise to increasingly complex, collective responses. Predicting these collective behaviors, however, remains largely restricted to steady-state characterization, as the dynamic coupling between time-varying flows and multiple interacting structures is difficult to model. In this paper, we investigate numerically the collective interaction of multiple asymmetric leaflets within a channel at low-Reynolds number. By utilizing symmetrically oscillating plates rather than a pressure-driven flow to isolate the system from background asymmetries, we characterize how these interacting structures generate a net fluid transport. We develop an analytical framework to evaluate transport in the steady limit, which we subsequently extend to account for time-dependent channel oscillations, providing a complete dynamic description of the coupled fluid-structure system. Our results demonstrate that high leaflet densities maximize collective interactions and net transport. Furthermore, we define an elastoviscous number comparing viscous hydrodynamic forces to the restorative elastic forces of the leaflets, and uncover an optimal value that maximizes the net flow. This framework establishes a foundation for analyzing how collective slender structures interact dynamically within viscous environments, laying the groundwork for future studies on flow control in biological fluid transport and microfluidic design.
\end{abstract}

%\begin{keywords}
%Authors should not enter keywords on the manuscript, as these must be chosen by the author during the online submission process and will then be added during the typesetting process (see \href{https://www.cambridge.org/core/journals/journal-of-fluid-mechanics/information/list-of-keywords}{Keyword PDF} for the full list).  Other classifications will be added at the same time.
%\end{keywords}

%{\bf MSC Codes }  {\it(Optional)} Please enter your MSC Codes here
\newpage
\section{Introduction}

%Across scales and flow dynamics, passive structures deforming in a flow have been used to control flow properties.
%OR: In nature, we have a lot of confined flow at low Reynolds (well microfluidic too), how do we embed control in these systems? We bring nonlinearity via this below.

Organisms utilize confined flows to transport matter and information across large scales and complex topologies \citep{moore_lymphatic_2018,lu2018biaxial,francis2009scaling}. At the low-Reynolds-number limit, these flows are regulated through the passive or active deformation of the confining boundaries themselves (vessels walls) or internal structures (leaflets, cilia). For passive soft boundaries, this regulation emerges from a two-way coupling: the boundaries deform in response to hydrodynamic stresses, while the resulting change in geometry simultaneously redefines the flow field \citep{levin2024asymmetric,park2021fluid}.

These observations have motivated extensive research in bio-inspired fluidic control. Single flexible structures have been shown to introduce nonlinear relationships between flow rate and pressure drop, with the nature of the nonlinearity depending on the geometry of the structure \citep{brandenbourger_tunable_2020,park2018viscous,10.1039/9781782628491,martinez2024fluidic}, and oscillatory flows further enriching this response through the dynamic deformation of the structure \citep{levin2024asymmetric}. Such local control on the flow have enabled a large body of work in microfluidics and soft robotics \citep{wehner2016integrated,mosadegh2010integrated}.

When multiple structures are arranged in arrays, their hydrodynamic interactions give rise to collective regimes in which the deformation of individual elements is set by the array as a whole \citep{thomazo_collective_2020}, and flow regulation emerges from the collective obstruction of the channel \citep{alvarado_nonlinear_2017,jambon-puillet_dense_2025}. The spatial distribution of structures has been shown to control flow nonlinearities across different configurations, whether driven by an external pressure gradient \citep{paludan2024elastohydrodynamic} or by active wall contractions \citep{winn2026unidirectional,winn2024operating,garcia2025spontaneous}. These studies demonstrate that fluid–structure interactions not only introduce mechanical nonlinearities but also break the inherent symmetries of the Stokes regime, frequently leading to the emergence of unidirectional transport and rectified flow. Despite these advances, existing studies have focused predominantly on single structures or steady-state regimes,  thereby neglecting the rich temporal dynamics that may emerge from the interaction of multiple structures and potentially give rise to more finely tuned transport regimes. 

In this work, we investigate how such dynamical effects influence the emergence and efficiency of flow rectification, using the canonical squeeze flow between two parallel plates as a model system. In the absence of soft structures, this configuration is governed by the reversible kinematics of the Stokes regime, yielding zero net transport regardless of the oscillation profile. By introducing compliant leaflets anchored to the bottom plate, we demonstrate how flow-induced reconfiguration breaks this symmetry, resulting in a rectified, unidirectional flow over a complete cycle. We investigate how the leaflets’ elasticity and density, and the channel geometry and dynamic oscillation influence flow directionality and transport across the channel.

To explore this parameter space, we employ fully coupled fluid-structure interaction simulations utilizing an immersed boundary method. This numerical approach is complemented by an analytical continuum-limit model that is extended phenomenologically to capture the system behavior across arbitrary leaflet densities. Our results demonstrate that collective interactions between the leaflets (high leaflet densities) maximize flow rectification and highlight an optimum ratio between the viscous hydrodynamic forces and the restorative elastic forces of the leaflets. While the resulting flow rectification is frequency-independent, we show that it is ultimately limited by the finite viscous relaxation timescale of the leaflets.

The remainder of this paper is organized as follows. Section 2 characterizes the mechanism of flow rectification emerging from the system and details the numerical methods used throughout this study. In Section 3, we develop an analytical framework to isolate the influence of the leaflets elasticity and density on flow rectification in the steady limit. This steady framework is subsequently extended to account for time-dependent channel oscillations in Section 4, providing a complete dynamic description of the system.

\section{Flow rectification in an oscillating fluidic channel}
\label{sec:results}

\begin{figure}[t!]
    \centering
    \includegraphics[width=0.99\textwidth]{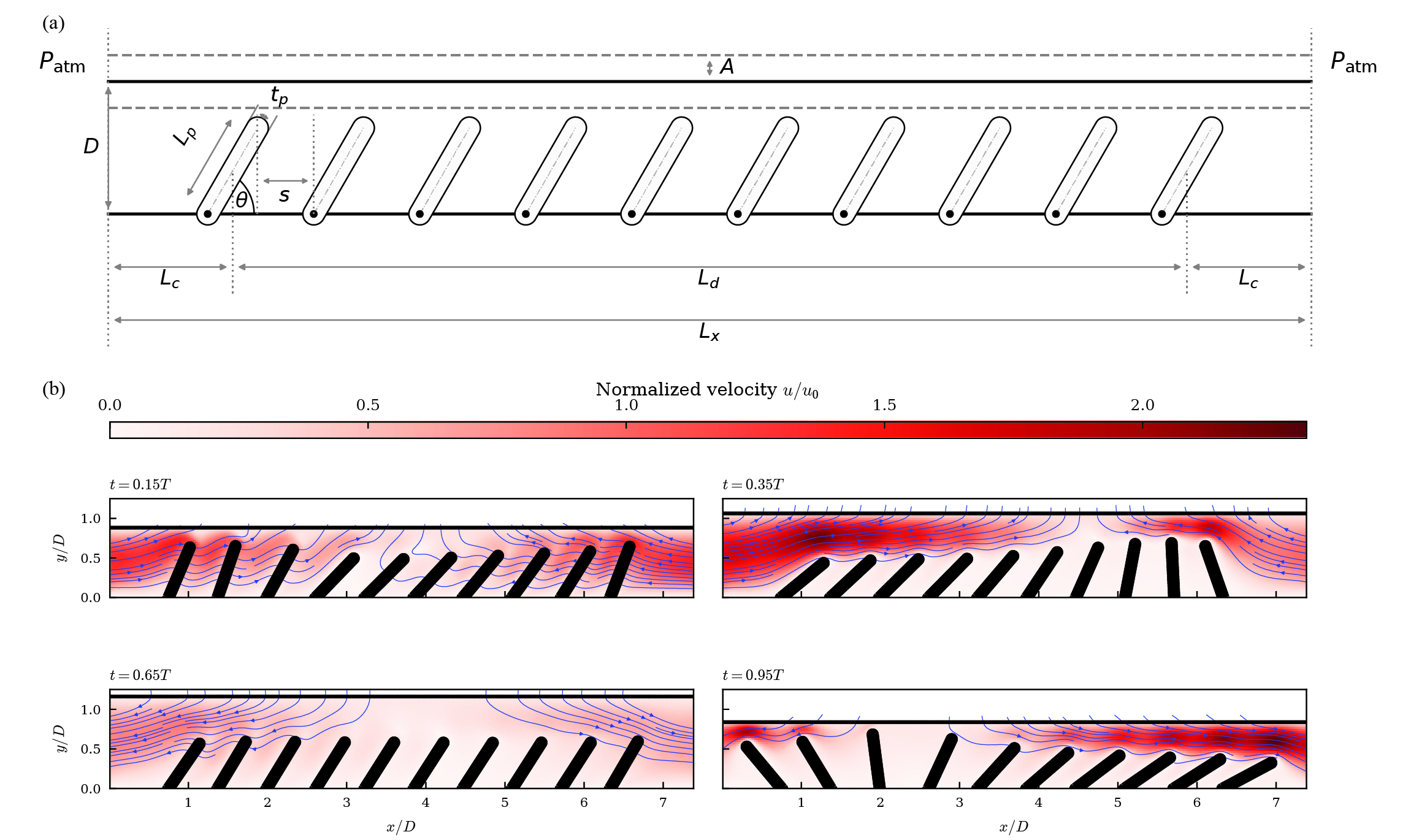}
    \caption{\textbf{(a)} Schematic of the 2D channel geometry. A bed of leaflets is distributed along the bottom wall, with each leaflet modelled as a rigid plate of length $L_p$, thickness $t_p$, and rest angle $\theta_0$, attached to a rotational spring of stiffness $k$. The flow of an incompressible Newtonian fluid (density $\rho$, viscosity $\mu$) is driven by the vertical oscillation of the top wall $y(t)=D -A\cos(\omega t)$, with both channel ends open to atmospheric pressure $p_{atm}$. \textbf{(b)} Instantaneous velocity fields and streamlines at four phases of the oscillation cycle ($t = 0.15T, 0.35T, 0.65T$, and $0.95T$) for a leaflet density of $\phi=0.6$ and $\eta = 0.02$.}
    \label{fig1}
\end{figure}

The physical system is represented by a minimal two-dimensional model of a channel with length $L_x$ and width $D$, as illustrated in Figure~\ref{fig1}. A bed of leaflets is distributed on the bottom wall of the channel. Each leaflet is modeled as a rigid plate of length $L_p$, thickness $t_p$, and rest angle $\theta_0$ mounted on a rotational spring of stiffness $K$. The top wall oscillates vertically according to:
\begin{equation}
y(t)=D -A\cos{\omega t}
\end{equation}

where $A$ is the oscillation amplitude and $\omega$ the oscillating pulsation. This wall motion drives the flow of an incompressible Newtonian fluid with density $\rho$ and viscosity $\mu$. The left and right channel boundaries are open to the atmosphere at pressure $p_{\text{atm}}$.\\

\begin{figure}[t!]
    \centering
    \includegraphics[width=0.99\textwidth]{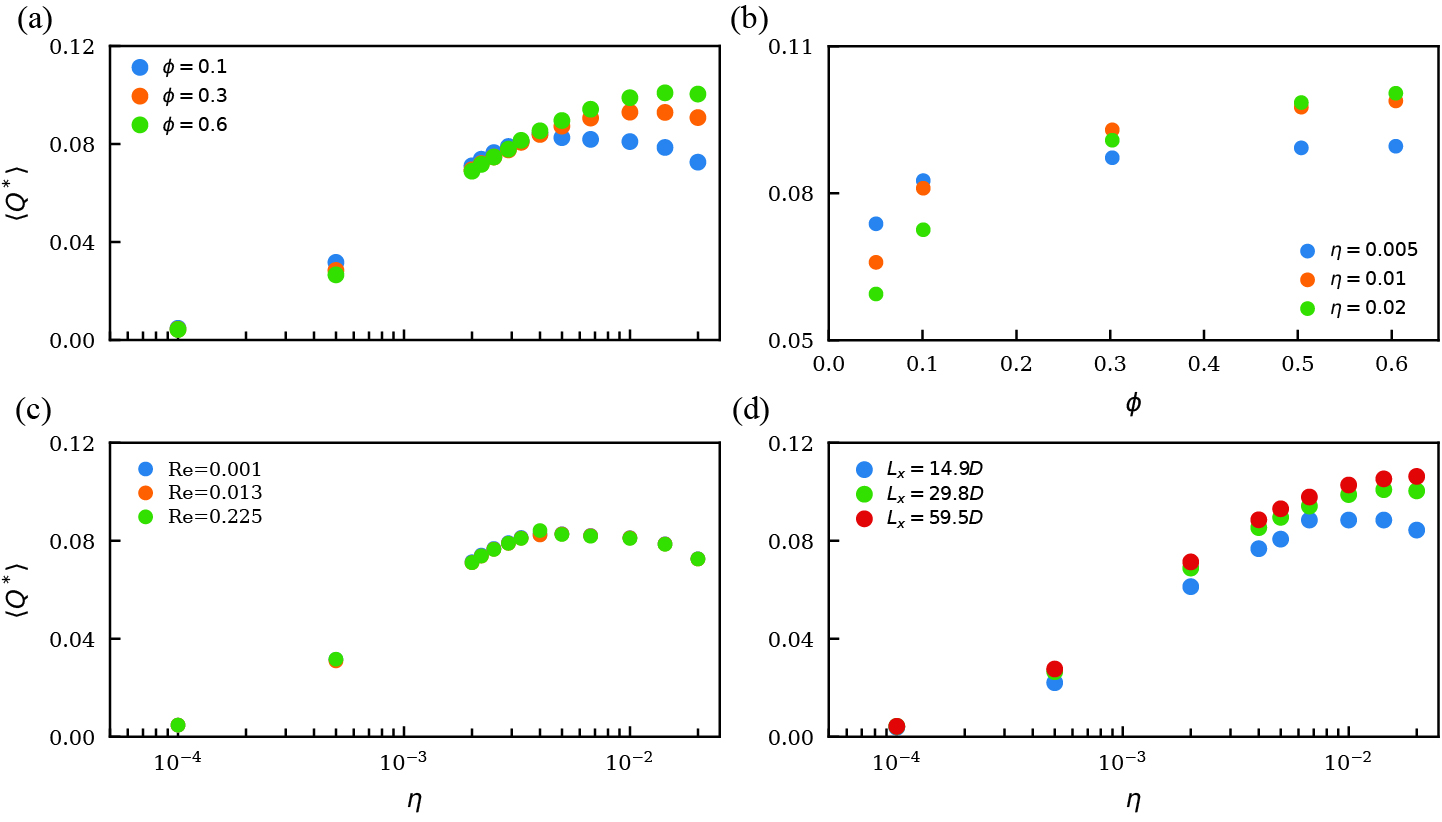}
    \caption{Characteristics of the flow rectification with varying system parameters. \textbf{(a)} Time averaged net flow rate $\langle Q^* \rangle$ as a function of the elasto-viscous number $\eta$ for three leaflet densities ($\phi = 0.1$ [blue], $0.3$ [orange], and $0.6$ [green]). Rigid leaflets ($\eta \ll 1$) produce negligible net flow, while an optimal flexibility exists that maximizes transport before decreasing for highly flexible leaflets. \textbf{(b)} Dependence of $\langle Q^* \rangle$ on the leaflet density $\phi = N L_p \cos(\theta_0) / L_x$, representing the projected area of the resting leaflets on the channel length where increasing density improves the flow directionality, plotted for three elasto-viscous numbers ($\eta = 0.005$ [blue], $0.01$ [orange], and $0.02$ [green]). \textbf{(c)} Invariance of $\langle Q^* \rangle$ with the Reynolds number ($Re = 0.001$ [blue], $0.013$ [orange], and $0.225$ [green]), confirming that transport remains independent of the oscillating frequency in the viscous regime. \textbf{(d)} Geometric effect, demonstrating the dependence of $\langle Q^* \rangle$ on the channel length ($L_x = 14.9D$ [blue], $29.8D$ [green], and $59.5D$ [red]).}
    \label{fig2}
\end{figure}

Figure~\ref{fig1}(b) presents a sequence of instantaneous flow states within the channel over a complete oscillation cycle. The details of the numerical method is described in section \ref{Numerical Method Overview}. As the top plate moves downward, an outward flow is induced toward both ends, causing the leaflets to deform. The gap narrows between the plate and the left leaflets, while it widens on the right. As the plate returns upward, the flow reverses direction and the leaflets deform accordingly. The asymmetry in the leaflet deformation is expected to generate flow rectification. Such rectification is quantified by the normalized net flow over one oscillation cycle:
\begin{equation}
    \langle Q^*\rangle = \frac{\langle Q^*_{in} \rangle+\langle Q^*_{out} \rangle}{2}
    \label{eq:flow_norm}
\end{equation}
where \(Q^*=Q/Q_{wall}\) and \(Q_{wall}=A \omega L_x\). A positive value indicates net forward transport. To understand the physical mechanisms governing this transport, we explore the parameter space by comparing fluid loading, leaflet stiffness, and geometry. We first define the elasto-viscous number $\eta$, which represents the ratio of viscous hydrodynamic forces to the restorative elastic forces of the leaflet:
\begin{equation}
    \eta = \frac{\mu\,U_0\,L_p^2}{K\,D},
\end{equation}
where $U_0=\frac{\omega A L_x}{2 D}$ is the characteristic out velocity in the case of zero leaflets. Large $\eta$ indicates a regime where the leaflets are easily reconfigured by the flow field. Figure~\ref{fig2}(a) shows that almost no net flow is created for rigid leaflets (small values of $\eta$). The net flow increases with $\eta$ until reaching a maximum, after which it decreases for very soft leaflets (large values of $\eta$). We also observe in Figure~\ref{fig2}(b) that the net flow depends on the density of leaflets defined as:
\begin{equation}
    \phi =  \frac{N L_p \cos{\theta_0}}{L_x},
\end{equation}
where $N$ is the number of leaflets and $\frac{L_p \cos{\theta_0}}{L_x}$ is the ratio between the projected area of the leaflets at its rest position and the length of the channel. The net flow rate increases with the leaflets density until saturating to a value that depends on the elasto-viscous number.

To characterize the role of the plate oscillation, we vary both the viscosity $\mu$ and the oscillation frequency $\omega$ while keeping $\eta$ constant by adjusting the stiffness of the leaflets $K$. The resulting net flow is shown in Figure~\ref{fig2}(c) as a function of the Reynolds number
\begin{equation}
Re=\frac{\rho\,U_0\,D}{\mu}.
\end{equation}
We limit this study to small Reynolds numbers and observe that the net flow rate does not depend on the viscosity $\mu$ and the oscillation frequency $\omega$ when $\eta$ is constant. 
This confirms that $\eta$ is the main non-dimensional number controlling the flow-structure behavior, regardless of the viscosity and oscillation frequency.
%\mb{The viscosity and driving frequency do not uniquely control the net transport; instead, their effects are entirely offset by the leaflet stiffness.} \sg{I'm not sure I get this last sentence}\mb{I let you rephrase}.  
Yet, in Figure~\ref{fig2}(d) we observe that the net flow rate depends on the channel length $L_x$. 

%For a constant elasto-visco number $\eta$, the net flow increases with the the aspect ratio of the channel $\frac{D^2}{A L_x}$. \mb{(Do you confirm that we kept L/D constant here?)}.\\

In summary, these results reveal that the net transport is governed by the visco-elastic number $\eta$, exhibiting a non-monotonic dependence with an optimal value for maximal flow (Figure~\ref{fig2}(a)). For a given elasto-viscous number, the net transport is independent of the oscillation frequency (Figure~\ref{fig2}(c)). Yet, the flow remains strongly coupled to the channel geometry (Figure~\ref{fig2}(d)). 
%In the following sections, we elucidate the physical mechanisms underlying these observations and develop a theoretical framework to describe the resulting transport. In Section \ref{Numerical Method Overview} we detail our numerical method. Section \ref{Steady Contracting Channels} considers a steady configuration with an imposed fluid flow through a stationary upper boundary.  We derive an analytical model for the flow rectification at the continuum limit and extend it to a limited number of leaflets. Section \ref{Time Dependant Effect}  considers the full dynamics of the plate oscillation extends the quasi-steady model to take into account the channel oscillation.\mb{This is very similar to the end of the intro, how should we deal with that?} \sg{Maybe we remove the last paragraph of the intro?}

%and demonstrates the origin of the geometric scaling via an analysis of the phase lag in the leaflets oscillation \sg{I don't think we do that anymore} \oa{Section \ref{Time Dependant Effect} considers the full dynamics of the top wall oscillation and demonstrates the origin the time dependant effect via an analysis of the phase lag in the leaflets oscillation}.

\section{Numerical Method}\label{Numerical Method Overview}

In this study, we simulate the fluid flow and its interaction with flexible leaflets using a lattice Boltzmann method (LBM) coupled with an immersed boundary method (IBM). 
To account for time-dependent effects while avoiding time-step restrictions inherent to LBM as an explicit solver, we employ a dual-time-stepping scheme accelerated with a multigrid approach. Only the essential numerical details are summarized below.\\

\subsection{Lattice Boltzmann Method}\

The flow is solved using a D2Q9 lattice-Boltzmann method with a TRT collision operator following \cite{gsell_lattice-boltzmann_2021}. The scheme recovers the incompressible Navier-Stokes equations in the low-Mach-number limit, which is ensured throughout the study by maintaining $U_0/ c_s < 0.05$.

The oscillation period is typically much larger than the hydrodynamic relaxation time. Solving the problem with a standard explicit LBM would therefore require a prohibitively large number of time steps. We consequently employ the dual-time-stepping (DTS) framework detailed in \cite{gsell_multigrid_2020}.
In brief, DTS decomposes the unsteady dynamics as a series of pseudo-steady problems, in which time derivatives of the flow quantities are treated as external source terms.
These unsteady terms are computed implicitly through an external second-order backward scheme.
Each pseudo-steady state is computed with a multigrid approach using the LB scheme as the core iterative scheme.
The time step is set to $T/20$, with $T$ the oscillation period. 
We checked that smaller time steps do not affect the leaflet and flow dynamics.
Each time step is solved through a series of multigrid V-cycles until a fixed convergence criterion is reached.

% \oa{The details of our DTS implementation are as follows: the physical time step is set to $1/20$ of the oscillation period, each time step is solved though a series of inner iterations—continuing until the convergence criterion is met—consisting of V-cycle multigrid}.

\subsection{Immersed-boundary method}

\begin{figure}[t]
    \centering
    \includegraphics[width=0.99\textwidth]{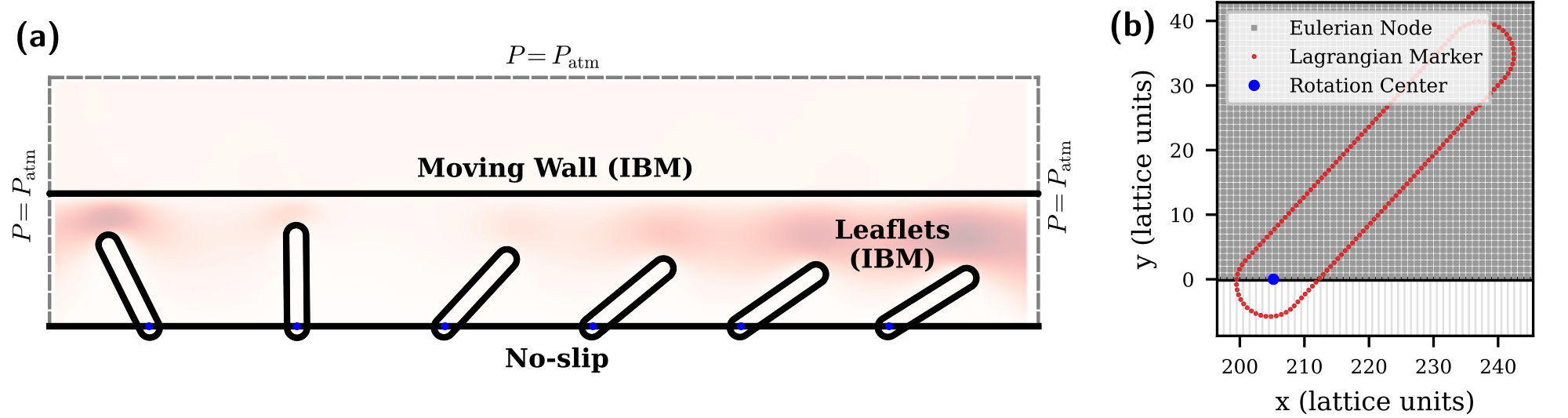}
    \caption{Numerical configuration. \textbf{(a)} Schematic of the two dimensional domain with the boundary conditions. The moving upper wall is modelled via an immersed boundary within a larger fixed Eulerian grid. A no-slip condition is enforced at the bottom wall, while constant atmospheric pressure ($P=P_{atm}$) is imposed at the left, right, and top boundaries. \textbf{(b)} Detailed view of the fluid structure of a single leaflet, illustrating the fixed Eulerian nodes for the LBM fluid solver, the discrete Lagrangian markers defining the leaflet geometry, with its fixed center of rotation.}
    \label{numerical_conf}
\end{figure}

The leaflet geometry is integrated into the flow through a set of Lagrangian markers with position $\bm{X_k}$ (Figure~\ref{numerical_conf}).
To ensure the fluid-structure coupling, i.e. the momentum exchange between the leaflets and the flow, we employ an immersed-boundary method.
At each time step, we first compute the flow velocity in absence of fluid-structure coupling, $\bm{u}^\dag$.
We then use an interpolation operator $\mathcal{I}$ to estimate the fluid velocity on each boundary marker,
\begin{equation}
  \mathcal{I}[\bm{u}^\dag](\bm{X_k}) = \sum_{\bm{x_i} \in \Sigma} \bm{u}^\dag (\bm{x_i}) \delta (\bm{x_i} -\bm{X_k}) \Delta S_i,
\end{equation}
where $\Sigma$ is the fluid domain, $\Delta S_i = \Delta x \Delta y = 1$ is the surface of one mesh cell and $\delta (\bm{x})$ is the interpolation kernel, decomposed as $\delta (\bm{x}) = \hat{\delta}(x) \hat{\delta}(y)$ with 
\begin{equation}
  \arraycolsep=1.4pt\def\arraystretch{2.}  
    \hat{\delta} (r) = \left\{
    \begin{array}{ll}
      \dfrac{1}{2d} \left( 1 + \cos \left( \dfrac{\pi r}{d} \right) \right), & ~~~ |r| \le d, \\
       0, & ~~~ |r| > d,
    \end{array}
          \right.
  \label{eqn:delta}        
\end{equation}
and $d$ is the radius of the kernel function, set to $3/2$ in the following.
We then compute the momentum correction on each marker as 
\begin{equation}
  \bm{F} (\bm{X_k}) = \dfrac{2 \rho_0}{\kappa \Delta t} \left(\bm{U_b}(\bm{X_k}) - \mathcal{I}[\bm{u}^\dag](\bm{X_k}) \right),
  \label{eqn:corrected_ibm}
\end{equation}
where $\bm{U_b}(\bm{X_k})$ is the boundary velocity prescribed by the leaflet motion.
In equation (\ref{eqn:corrected_ibm}), we use a kernel-dependent correction factor $\kappa$ preventing the boundary slip error classically arising in IBM \citep{gsell_direct-forcing_2021}.
With the present kernel function (\ref{eqn:delta}), $\kappa = 1/2$. 
Finally, the momentum correction (\ref{eqn:corrected_ibm}) is spread to the fluid nodes following 
\begin{equation}
  \bm{f}(\bm{x}) = \sum_{\bm{X_k} \in \Gamma} \bm{F} (\bm{X_k}) \delta (\bm{x} - \bm{X_k}) \Delta l_k,
\end{equation}
where $\Gamma$ denotes the immersed boundary and $\Delta l_k$ is the local distance between two boundary markers.
The momentum correction $\bm{f}(\bm{x})$ is treated as an external body force in the Navier-Stokes equations, that we integrate in the LBM framework following the scheme proposed by \cite{guo_discrete_2002}.
In particular, the fluid momentum is re-written as 
\begin{equation}
  \rho \bm{u} = \sum_{l=0}^N f_l \bm{c_l} + \dfrac{1}{2} \bm{f}.
\end{equation}

\subsection{Leaflet dynamics}

We model the leaflets as rigid structures rotating around their center of rotation (Figure~\ref{numerical_conf}(b)) as a response to the torque exerted by the fluid.
The quantity $\bm{F} (\bm{X_k})$ (\ref{eqn:corrected_ibm}) represents the force per unit boundary length required to enforce the no-slip condition and therefore corresponds to the hydrodynamic load exerted by the leaflet on the fluid.
We use it to compute the fluid torque on each leaflet as
%The local momentum correction $\bm{F} (\bm{X_k})$ (\ref{eqn:corrected_ibm}) accounts for the force exerted by one surface element to the surrounding fluid.
%We thus compute the fluid torque on each leaflet as 
\begin{equation}
  T = - \sum_{\bm{X_k} \in \Gamma} \left( \bm{X_k} - \bm{A} \right) \times \bm{F} (\bm{X_k}) \Delta l_k, 
  \label{eqn:torque}
\end{equation}
where $\bm{A}$ is the center of rotation of the leaflet and $\times$ denotes the cross product.
Since the leaflets are modeled as rigid plates attached to torsional springs, their rotational dynamics follow the balance between hydrodynamic, elastic, and inertial torques: 
\begin{equation}
    J \ddot{\Delta \theta} = T - K \Delta \theta,
    \label{eqn:leaflet}
\end{equation}
where $K$ is the torsional stiffness, $\Delta \theta$ is the angular deflection, $J$ is the moment of inertia of the leaflet given by $J = \frac{1}{3} m L_p^2$, and the leaflet mass $m$---per unit length---is approximated by $m = \rho_s L_p t_p$ with $\rho_s = \rho$ the density and $t_p$ the leaflet thickness.
In all the simulations presented in the following, the inertial term in equation (\ref{eqn:leaflet}) remains negligible in practice, i.e. $T \approx K \Delta \theta$ at each time step.

We treat the leafleat dynamics implicitly by integrating equation (\ref{eqn:leaflet}) together with the fluid dynamics through the DTS scheme.
The inertial term is discretized though a second-order backward scheme and we update the leaflet position and velocity using a series of inner iterations.
At each inner iteration, the flow variables are updated through a multigrid cycle, the new fluid torque (\ref{eqn:torque}) is computed and the leaflet position and velocity are updated accordingly.
This loop is repeated until convergence of the pseudo-steady problem is reached.

\subsection{Numerical configuration and boundary conditions}

The computational domain consists of a 2D rectangle discretized with a Cartesian and uniform mesh.
To avoid boundary condition issues with the upper plate motion, we model it with an immersed boundary and include it in a bigger domain with fixed height (Figure~\ref{numerical_conf}).
The number of LB nodes between the bottom wall and the mean upper-plate position is fixed to $\approx 70$.
In the horizontal direction, the number of nodes $N_x$ is set such that $\Delta x = \Delta y$ and varies with the domain length $L$.
We apply a no-slip condition on the bottom wall using a bounce-back scheme \citep{kruger_lattice_2017}.
On the left, right and top boundaries, a constant pressure is imposed using an anti-bounce-back scheme.
Constant-pressure boundaries reproduce the experimental situation of a channel connected to large reservoirs and allow the rectified flow rate to emerge naturally from the fluid-structure interaction.
All the simulations are initialized with the fluid and the leaflets at rest. In time-dependent simulations, we apply a series of oscillation cycles to ensure that a steady regime is reached.

%After having presented the essential results in the oscillating case in section \ref{sec:results}, we next focus on a steady configuration in section \ref{Steady Contracting Channels}, to facilitate the analysis of the flow rectification.
%We then further analyze the time-dependent effects in section \ref{Time Dependant Effect}.

In the following, the numerical framework described above is first applied to a simplified steady forcing configuration (section \ref{Steady Contracting Channels}), which allows the mechanisms governing flow rectification to be isolated before returning to the fully oscillatory problem (section \ref{Time Dependant Effect}).

\section{Flow rectification in the steady limit}
\label{Steady Contracting Channels}

\subsection{Simulations in the steady limit}

\begin{figure}[t!]
    \centering
    \includegraphics[width=0.99\textwidth]{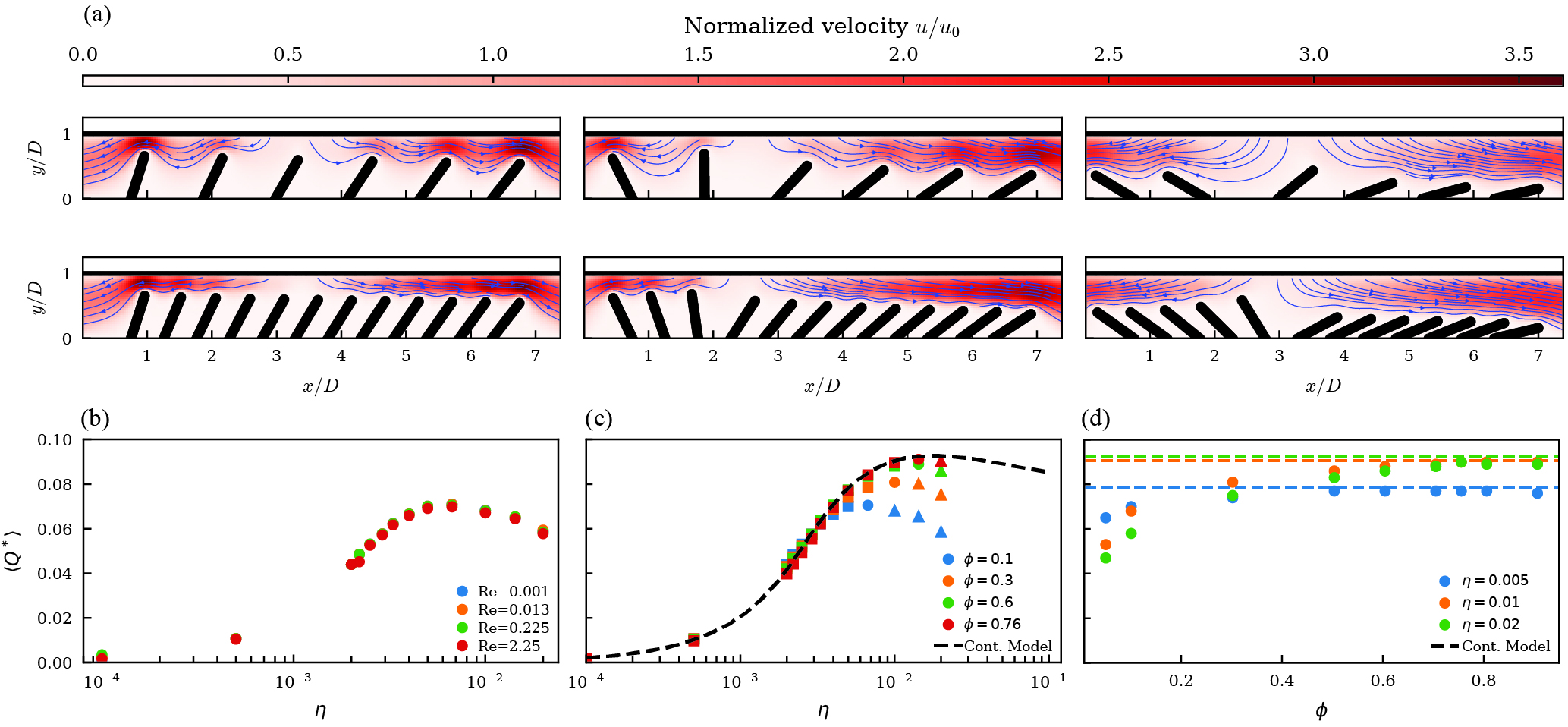}
    \caption{Flow rectification characteristics in the steady limit. To isolate structural and geometric effects from the dynamics of the plate oscillation, the moving boundary is replaced by a fixed wall at height $D$ subjected to a uniform normal velocity $U_{wall}$ mimicking channel contraction and expansion. \textbf{(a)} Flow states across varying elasto-viscous numbers ($\eta = 0.002, 0.0142$, and $0.1$, from left to right) and leaflet densities ($\phi = 0.3$ and $0.6$, top and bottom rows). \textbf{(b)} Invariance of the net flow rate $\langle Q^* \rangle$ with the Reynolds number ($Re = 0.001$ [blue], $0.013$ [orange], $0.225$ [green], and $2.25$ [red]) in the viscous regime. \textbf{(c)} Dependence of $\langle Q^* \rangle$ on the elasto-viscous number $\eta$ for four leaflet densities ($\phi = 0.1$ [blue], $0.3$ [orange],$0.6$ [green], and $0.76$ [red]) demonstrating that the non-monotonic trend and the optimal flexibility (dot) persist under the steady forcing. \textbf{(d)} Dependence of $\langle Q^* \rangle$ on the leaflet density $\phi$ at different values of $\eta$ where increasing density for softer leaflets improves the flow directionality, plotted for three elasto-viscous numbers ($\eta = 0.005$ [blue], $0.01$ [orange], and $0.02$ [green]). In panels \textbf{(c)} and \textbf{(d)} the discrete simulation results are compared against theoretical predictions derived from a continuous limit model ($\phi \rightarrow \infty$). }
    \label{fig3}
\end{figure}

To isolate the effects of geometry and elasticity from the effect of the plate oscillation, we first simplify the system to a steady case in which the top wall is fixed, with the distance between the two walls held at $D$. We impose a constant negative (positive) vertical flow velocity $U_{wall}$ to mimic contraction (relaxation) of the channel. Figure~\ref{fig3}(a) show different flow states for selected values of elasto-viscous numbers and leaflet densities. For each studied configuration we define the net flow $\langle Q^*\rangle\ $ by averaging the flows produced by one steady contraction and one steady relaxation. As shown in Figure~\ref{fig3}(b), the net flow remains invariant with respect to the Reynolds number in the steady regime. The system response is characterized solely by the interplay between the elasto-viscous number $\eta$ and the structural density $\phi$.

The steady regime recovers the non-monotonic dependence on $\eta$ previously observed in the unsteady case. While the specific magnitudes differ, the persistence of a distinct maximum suggests that the optimal transport mechanism is fundamentally rooted in the steady fluid-structure interaction. To rationalize this observation, we build a model considering a continuous limit (high leaflet density $\phi \rightarrow \infty$) for which the discrete leaflets are represented by a continuous profile.% and where the distance between the leaflet tips and the top wall is defined as $h(x)$.

\subsection{Continuum torque model}

\begin{figure}[t!]
    \centering
    \includegraphics[width=0.99\textwidth]{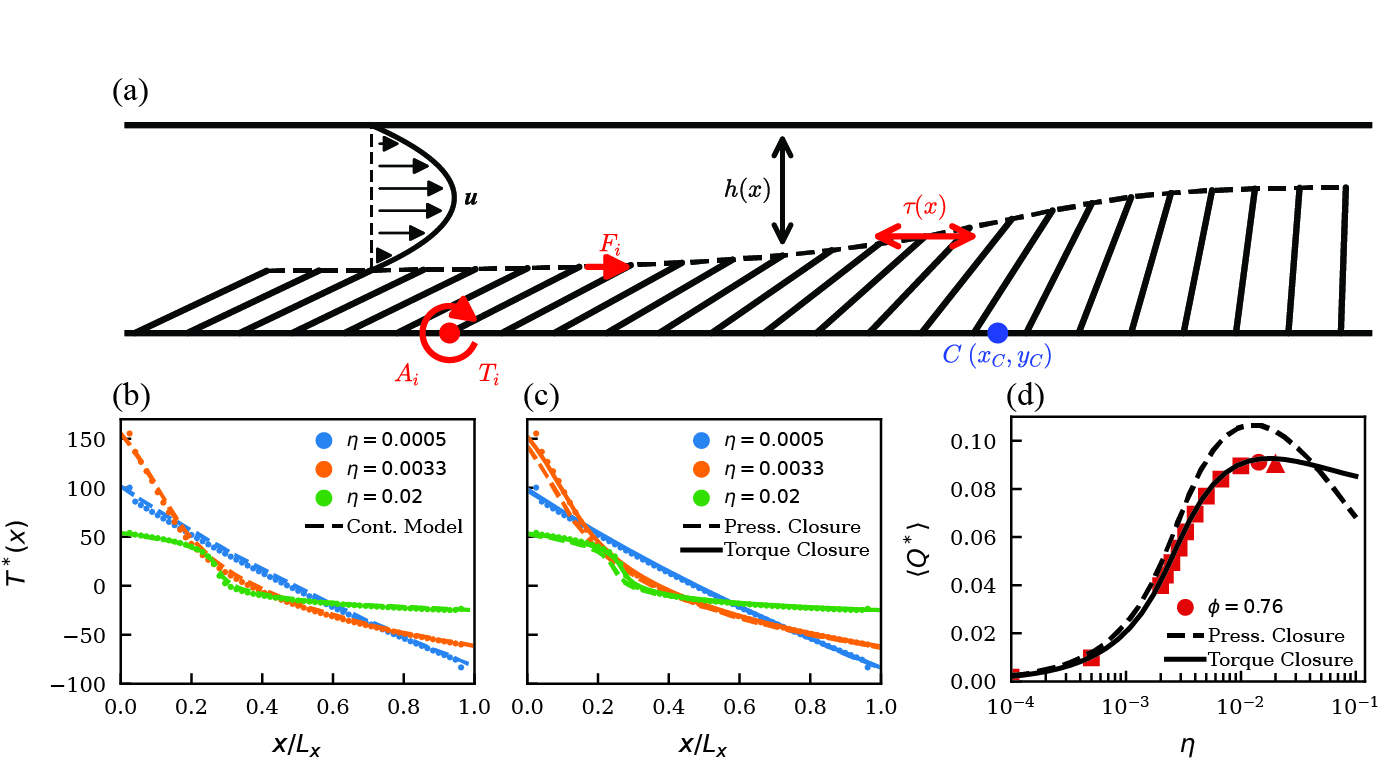}
    \caption{Theoretical formulation of the continuum flow rectification model.$(a)$ Schematic of the continuum approximation, defining the local gap height $h(x)$ and the variables required to establish the local torque balance and global angular momentum conservation. $(b)$ Validation of the theoretical torque distribution $T(x)$ (equation \ref{eq:momentum_balance_final}) against the simulation data for three elasto-viscous numbers ($\eta = 0.0005$ [blue], $0.0033$ [orange], and $0.02$ [green]) at a fixed leaflet density $\phi=0.76$ using the stagnation points $x_s$ from the simulations. $(c)$ Comparison of the predicted torque profiles for the same three values of $\eta$ obtained using two macroscopic closures: a global pressure constraint ($P(0)=P(L_x)$) versus a zero net torque condition derived from angular momentum conservation ($\int_0^{L_x} T(x)\,dx = 0$). $(d)$ Resulting predictions for the net flow rate $\langle Q^* \rangle$. The angular momentum closure yields significantly improved quantitative agreement with high density simulation data, resolving the over prediction from the pressure closure in the soft regime (high $\eta$).}
    \label{fig:twomodel}
\end{figure}

In the continuum limit, we define $h(x)$ as the distance between the leaflet tips and the top wall (Figure~\ref{fig:twomodel}(a)).
We start by considering the continuity equation for an incompressible fluid ($\nabla \cdot \mathbf{u} = 0$). The change in flow rate $Q(x)$ along the channel equals the volume displaced by the wall:
\begin{equation}
    \frac{dQ}{dx} = U_{wall}.
\end{equation}
We assume the existence of a stagnation point $x_s$ where the net flow vanishes ($Q(x_s)=0$), dividing the channel into two regions where flow exits from the left and right boundaries. Integrating from $x_s$ to any position $x$ gives
\begin{equation}
    Q(x) = \int_{x_s}^x U_{wall} \, dx = U_{\text{wall}}(x_s - x)
    \label{eq:mass_conservation_derived}
\end{equation}
The term $(x_s - x)$ represents the cumulative length of the channel contributing to the flow at position $x$. The flow rate increases linearly with distance from the stagnation point.

In the lubrication limit ($Re \ll 1$, $D/L \ll 1$), the pressure-driven flow through the local gap $h(x)$ follows
\begin{equation}
    Q(x) \propto \frac{\partial P}{\partial x}(x) \frac{h(x)^3}{\mu},
    \label{eq:mass_conservation_lubrication}
\end{equation}
and the local shear force---per unit length---at the leaflet tips is proportional to the local pressure gradient and the gap height:
\begin{equation}
    \tau(x) \propto \frac{\partial P}{\partial x}(x) h(x).
    \label{eq:shear_tips}
\end{equation}

We assume that the hydrodynamic load is concentrated at the leaflet tip. 
The fluid torque applied to the tip $T(x)$ is modeled as the product of the local shear force $\tau(x)$ and the effective lever arm and is balanced by the elastic restoring torque from the rotational spring:
\begin{equation}
    T(x) \approx \underbrace{ \tau(x) }_{\text{Force}} \cdot \underbrace{ L_p \sin\theta(x) }_{\text{Lever Arm}} = K\bigl(\theta(x) - \theta_0\bigr),
    \label{eq:momentum_balance_1}
\end{equation}
where the effective lever arm is the vertical projection of the leaflet length---$L_p \sin\theta$---from the pivot. 
The shear force in the lubrication limit scales as $\tau(x) \propto Q(x) / h(x)^2$. Substituting this into the torque balance (\ref{eq:momentum_balance_1}) yields the expression for the torque locally exerted on the leaflets:
\begin{equation}
T(x) = \frac{\alpha \mu\,U_{\text{wall}} (x_s - x) L_p \sin\theta(x)}{\bigl(D - L_p\sin\theta(x)\bigr)^2} = K\bigl(\theta(x) - \theta_0\bigr),
\label{eq:momentum_balance_final}
\end{equation}
where the prefactor $\alpha$ is a fitting parameter.
In Figure~\ref{fig:twomodel}(b), we extract the value of $x_s$ from simulation data and show that expression (\ref{eq:momentum_balance_final}) accurately represents the torque distribution for various values of $\eta$ and for $\phi=0.76$.
Here, we fixed $\alpha = 660$, which provides a good fit in all the cases shown in Figure~\ref{fig:twomodel}(b).

\subsection{Model closure and flow prediction}

To fully close the torque model (\ref{eq:momentum_balance_final}), we need an extra equation fixing the value of $x_s$.
We first propose to use the left and right pressure conditions, $P(x=0) = P(x=L_x)$, which implies 
\begin{equation}
  \int_{0}^{L_x} \dfrac{\partial P}{\partial x} dx = 0.
\end{equation}
Noting that in our lubrication model $\partial P / \partial x = Q(x) \mu / h^3(x) = U_{wall} \mu (x_s - x) / h^3(x)$, this pressure condition results in 
\begin{equation}
  x_s = \dfrac{\int_0^{L_x} x/h^3(x) dx}{\int_0^{L_x} 1/h^3(x)dx}.
  \label{eq:x_s_1}
\end{equation}
Substituting $x_s$ in equation (\ref{eq:momentum_balance_final}) and numerically solving for the torque distribution results in a reasonable prediction of $T(x)$ in various conditions (Figure~\ref{fig:twomodel}(c)).
The value of $x_s$ also allows us to compute the net flow rate,
\begin{equation}
 Q =\frac{U_{\text{wall}}(2x_s - L_x)}{2}.
\end{equation}
As shown in Figure~\ref{fig:twomodel}(d), the predicted flow rate is in qualitative agreement with the simulations performed at high leaflet densities.
However, quantitative discrepancies are notable, especially in the high $\eta$ region.

We propose an alternative closure model based on torque balance in the system.
At steady state, the conservation of angular momentum in a control fluid volume reads
\begin{equation}
 \bm{T_{tot}^{(C)}} = \int_S \rho \bm{r^{(C)}} \times \bm{u} (\bm{u} \cdot \bm{n}) dS,
 \label{eqn:ang_mom}
\end{equation}
where $\bm{T_{tot}^{(C)}}$ is the total torque applied to the fluid volume by the leaflets, expressed at a reference point $C$ with coordinates $(x_C, y_C)$.
The right-hand side of equation (\ref{eqn:ang_mom}) accounts for the angular momentum fluxes at the domain boundaries---with $r^{(C)}$ the position vector with respect to $C$--- that we assume to vanish here, i.e. $\bm{T_{tot}^{(C)}} = \bm{0}$.
For a series of $M$ leaflets subjected to tip forces $(\bm{F_1}, \bm{F_2}, ... \bm{F_M})$ generating torques $(\bm{T_1}, \bm{T_2},..., \bm{T_M})$ at their axis of rotation $A_i$ (Figure~\ref{fig:twomodel}(a)), the total torque exerted to the fluid volume at point $C$ is 
\begin{equation}
  \bm{T_{tot}^{(C)}} = \sum_{i=1}^M - \bm{T_i} - \sum_{i=1}^M \bm{r}^{(C)}(A_i) \times \bm{F_i},
\end{equation}
where $\bm{r}^{(C)}(A_i)$ denotes the vector between $C$ and $A_i$.
Assuming the tip forces $\bm{F_i}$---assimilated to the shear stress $\tau(x)$ in the continuum limit---to be horizontal and going back to a continuous formulation, the total torque projected along the torque direction becomes 
\begin{equation}
  T^{(C)}_{tot} = \int_0^{L_x} - T(x) dx - y_c \int_0^{L_x} \tau(x)dx.
\end{equation} 
Recalling that $T^{(C)}_{tot} = 0$ and assuming $\int_0^L \tau(x)dx = 0$---i.e. no momentum fluxes in the horizontal direction, we obtain the condition 
\begin{equation}
  \int_0^{L_x}  T(x) dx = 0.
\end{equation}
This condition, applied as a closure model to (\ref{eq:momentum_balance_final}), improves the prediction of the torque profile $T(x)$ (Figure~\ref{fig:twomodel}(c)) and results in an excellent prediction of the directional flow in the high-$\phi$ regime (Figure~\ref{fig:twomodel}(d)).

%\subsection{Leaflet deformation and flow rectification regimes}
\subsection{Influence of the leaflets density}

%\mb{( This might appear earlier depending on how we fit above )} \oa{the simulations here, we're limited with elasticity it would be difficult to show the elastic regime but probably the optimum and the stiff regime, or can we do it for lower density for the simulations ?}

\begin{figure}[t!]
    \centering
    \includegraphics[width=0.99\textwidth]{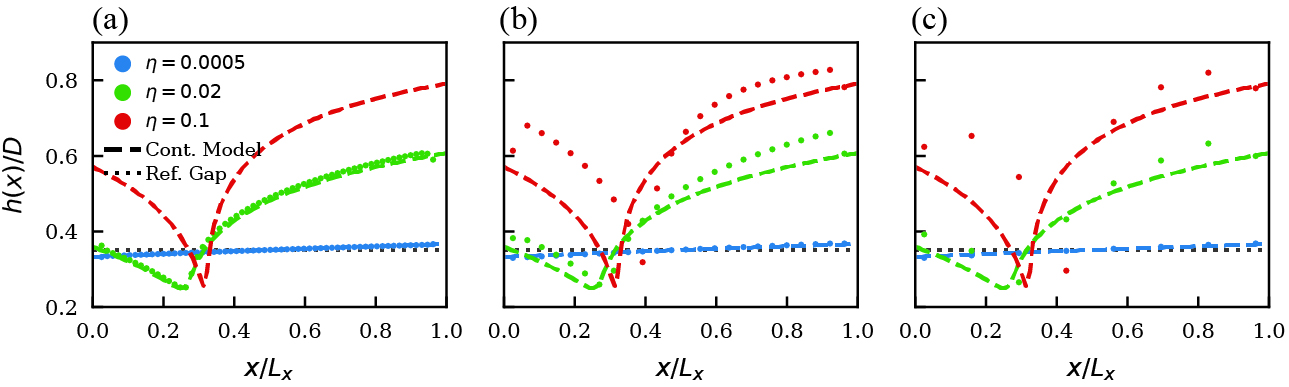}
    \caption{Gap profiles $h(x)$ illustrating the effect of leaflet stiffness on the flow rectification within three leaflet densities: \textbf{(a)} $\phi = 0.1$, \textbf{(b)} $\phi = 0.3$, and \textbf{(c)} $\phi = 0.76$. In each panel, the analytical continuous envelope is compared to simulations. The profiles highlight three distinct flexibility regimes: stiff leaflets ($\eta = 0.0005$ [blue]) yielding a monotonically decreasing gap; optimally flexible leaflets ($\eta = 0.02$ [green]) that lead to deforming past $90^\circ$ and create a non-monotonic profile that shifts the flow constriction upstream; and very soft leaflets ($\eta = 0.4$ [red]) that reduces the geometric asymmetry. Within all regimes, decreasing the density $\phi$ results in larger individual leaflet deformation due to non-presence of inner leaflets and shifting from the theoretical model approximation.}
    \label{fig4}
\end{figure}

We have now established that the flow in the continuum limit is governed by the continuous envelope of the leaflet height distribution. Therefore, the increase of flow rate from small to large density observed in Figure~\ref{fig3}(c, and d) reflects finite-size effects. As $\phi$ increases, the spatial gaps between discrete leaflets diminish, and the effective boundary profile asymptotically converges toward the continuum limit.
%Fig.~\ref{fig3}cd shows that the net flow rate increases with the leaflet density $\phi$. Since the flow in the continuum limit is governed by the continuous envelope of the leaflet height distribution (Eq. \ref{eq:x_s_1}), the density dependence reflects finite-size effects. As $\phi$ increases, the spatial gaps between discrete leaflets diminish, and the effective boundary profile asymptotically converges toward the continuum limit. 
However, Figure~\ref{fig3}(c, and d) reveals a more nuanced behavior: the scaling of the net flow with density is also modulated by the elasto-viscous number $\eta$. To further investigate this aspect, we show in Figure~\ref{fig4} the height of the channel h(x) for the continuum model (dashed lines) and 3 different densities, $\phi=0.1, 0.3,$ and $0.76$. Both the model and the data consistently exhibit three distinct regimes.

% physical mechanism driving the transition between the density insensitive (stiff) and collective (soft) regimes. The behavior is governed by the streamwise profile of the channel gap height $h(x)$. Figure~\ref{fig:gap_theory} presents the theoretical prediction of this profile and its asymmetry during the contraction phase.

For stiff leaflets (e.g. $\eta = 0.0005$), the deformation is small, and the maximum leaflet angle remains below $90^\circ$. In this regime (Figure~\ref{fig4} blue lines and dots), the gap $h(x)$ decreases monotonically along the channel from right to left. Further increase in $\eta$ enhances the leaflet deformation and the resulting spatial asymmetry, driving a larger net flow (Figure~\ref{fig3}(a) and Figure~\ref{fig3}(c) squares).

%\mb{( I'm not sure about the interpretation below, does it bring something quantitative? Because hydrodynamic resistance in the lubrication regime scales strongly with the inverse cube of the gap height ($1/h^3$), the absolute narrowest point controls the majority of the total resistance. In the monotonic case, this minimum gap occurs strictly at the leading leaflet (the left outlet). This single leaflet acts as a localized choke point. Also because the resistance is not distributed, this can explain why adding less/more leaflets has negligible effect.)}

For softer leaflets (e.g. $\eta = 0.02$), the leading leaflets deform past $90^\circ$ (Figure~\ref{fig4}(a) green line and dots). This creates a non-monotonic gap profile $h(x)$, effectively shifting the flow constriction upstream away from the outlet. This leads to a sub-linear increase in net flow, up to a maximum (Figure~\ref{fig3}(a) and Figure~\ref{fig3}(c) dot). 

Further increases in elasticity (e.g., $\eta=0.4$) lead to excessive deformation (Figure~\ref{fig4}(a) red line and dots). The leaflets flatten excessively across the entire domain, which reduces the asymmetry and leads to a steady decline as $\eta$ increases further (Figure~\ref{fig3}(a) and Figure~\ref{fig3}(c) triangles).

For every regime, we observe that lower densities produce a larger deformation of each individual leaflet. This is likely explained by the increasingly more complex flow field that penetrates the inter-leaflet gaps (Figure~\ref{fig3}(a). At lower-densities, leaflets experience higher hydrodynamic loading across their full length, leading to greater deformation. This confirms that the scaling of the net flow with density is also modulated by the elasto-viscous number.
%Yet, the flow field becomes increasingly complex as the fluid penetrates the inter-leaflet gaps (Fig. \ref{fig3}abc). This increasing flow complexity likely explains why the leaflet density $\phi$ modulates the transition threshold between regimes. At lower-densities, leaflets experience higher hydrodynamic loading across their full length, leading to greater compliance. As a consequence, the critical visco-elastic number $\eta$ required to reach the maximum net flow decreases as the system becomes less dense.

\begin{figure}[t!]
    \centering
    \includegraphics[width=0.99\textwidth]{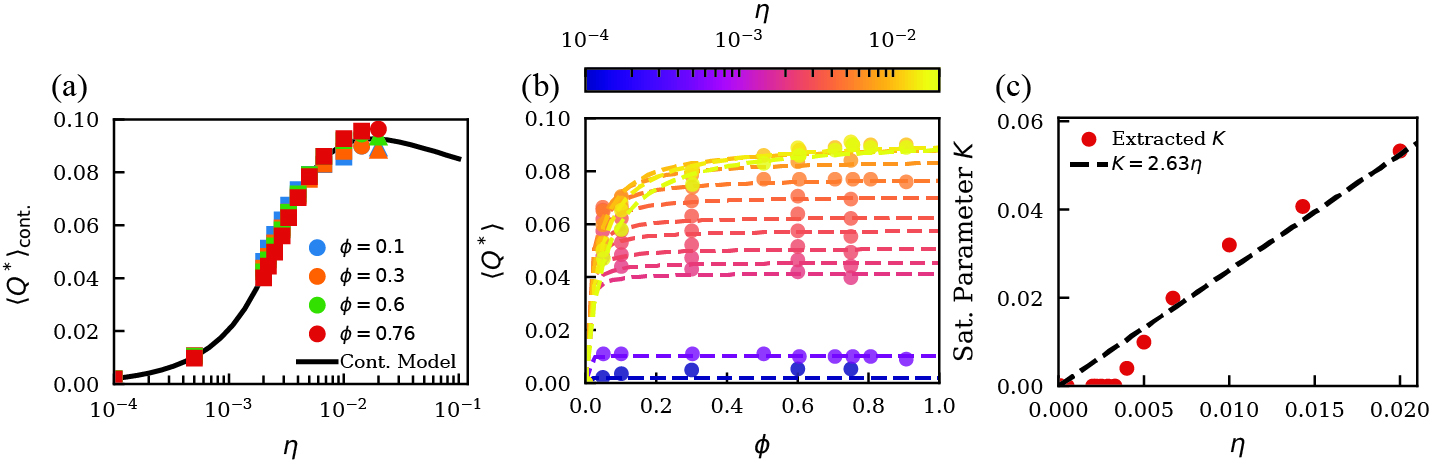}
    \caption{Phenomenological modelling of finite leaflet densities. \textbf{(a)} Collapse of the simulation data for four leaflet densities ($\phi = 0.1$ [blue], $0.3$ [orange],$0.6$ [green], and $0.76$ [red]) onto a master curve, validating the proposed phenomenological scaling across the elasto-viscous numbers $\eta$. \textbf{(b)} Steady net flow rate $\langle Q^* \rangle$ as a function of the leaflet density $\phi$. The simulations are described by the saturating phenomenological function $Q(\eta,\phi)=Q_{\rm continuum}(\eta) \phi / (\phi+K(\eta))$, which demonstrates that the flow asymptotically approaches the continuous limit as spatial gaps vanish. \textbf{(c)} Effect of the saturation parameter $K$ extracted from the individual fits in panel \textbf{(d)} as a function of $\eta$. The parameter $K$ is fitted with a linear dependence, which is used to plot the global collapse presented in panel \textbf{(a)}. }
    \label{pheno_model}
\end{figure}

These observations are also corroborated by Figure~\ref{fig3}(d), which shows that the net flow rate increases with the density as the effective boundary profile asymptotically converges toward the continuum limit. Yet, the rate at which this increase occurs depends on the elasto-viscous number. For leaflets that easily deform with the flow (large $\eta$),  optimum net flow (optimum deformation) is reached at small values of $\phi$, indicating that much larger $\phi$ is required to reach the continuum limit. The dependence of the net flow on density can be captured using a phenomenological saturating function:
\begin{equation}
Q(\eta,\phi)=Q_{continuum}(\eta) \frac{\phi}{\phi+K(\eta)},
\end{equation}\label{eq_density}
where $K(\eta)$ is any function of $\eta$. Figure~\ref{pheno_model}(b) shows the net flow as a function of leaflet density. While Figure~\ref{fig3} displayed only a representative subset of the data for visual clarity, here we present the full set of simulated cases to validate the model. We first fit Equation \ref{eq_density} using $K$ as a fitting parameter (dashed lines in Figure~\ref{pheno_model}(b)), then plot $K$ as a function of $\eta$ in Figure~\ref{pheno_model}(c). The function $K$ appears to increase linearly with $\eta$, except at small values of $\eta$, where we cannot access small values of $\phi$, which limits the reliability of the fit in this regime. Using the linear fit shown in Figure~\ref{pheno_model}(c), we rescale the data in Figure~\ref{pheno_model}, which collapse well onto a single master curve.

\section{Time-dependent dynamics}\label{Time Dependant Effect}

\begin{figure}[t!]
    \centering
    \includegraphics[width=0.99\textwidth]{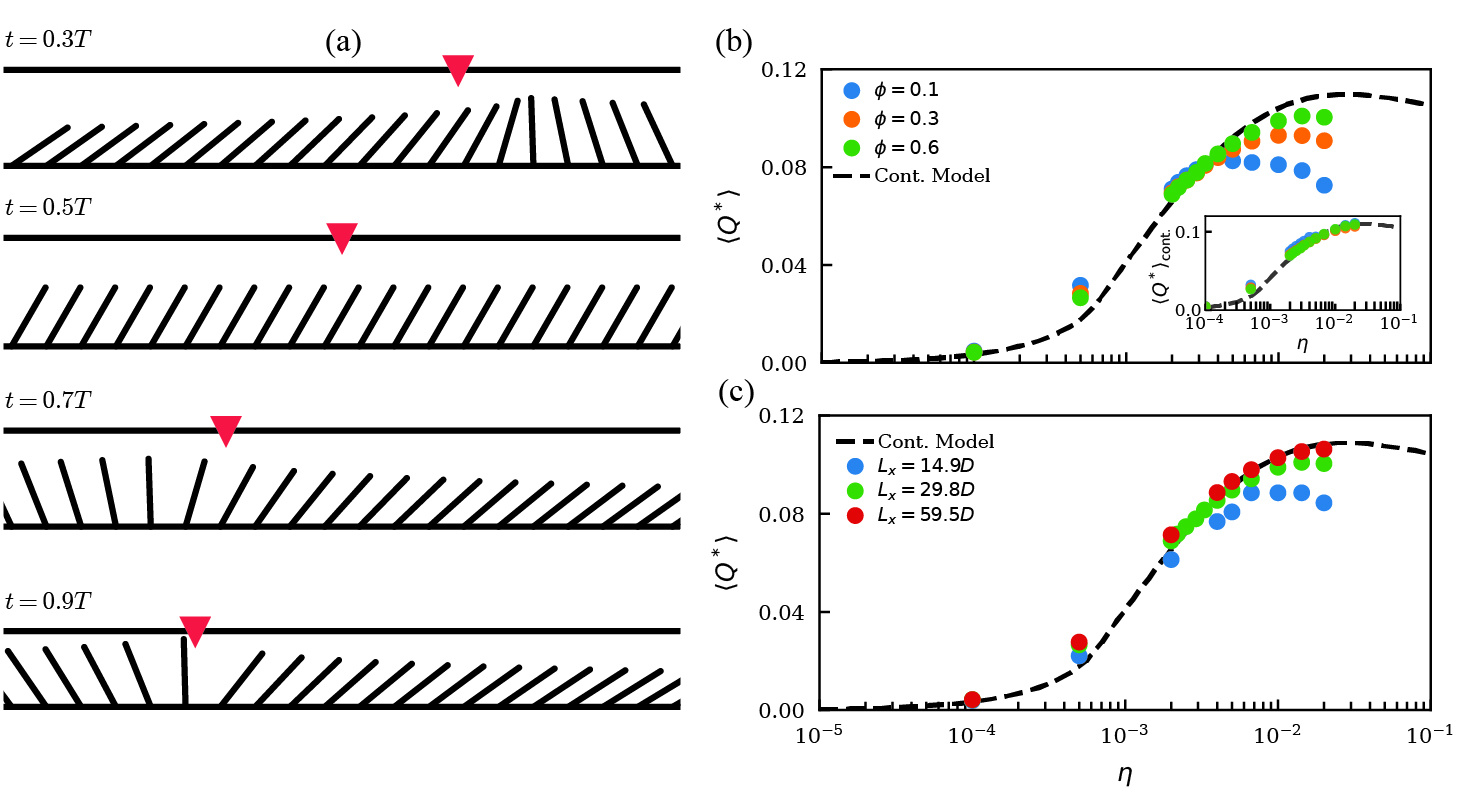}
    \caption{Extension of the continuum model to the oscillatory regime via a quasi-steady approximation. \textbf{(a)} Instantaneous configurations of the system at four phases of the oscillation cycle ($t = 0.3T, 0.5T, 0.7T$, and $0.9T$). \textbf{(b)} Comparison of the time averaged net flow rate $\langle Q^* \rangle$ as a function of the elasto-viscous number $\eta$ between the quasi-steady model and the simulations for three leaflet densities ($\phi = 0.1$ [blue], $0.3$ [orange], and $0.6$ [green]). The inset demonstrates that the phenomenological density scaling introduced in the steady limit, $Q(\eta,\phi) \sim \phi/(\phi+K \eta)$, successfully collapses the oscillatory data. \textbf{(c)} Geometric effect of the channel length ($L_x = 14.9D$ [blue], $29.8D$ [green], and $59.5D$ [red]) on the rectified flow where the quasi-steady model does not capture that effect. }
    \label{fig5}
\end{figure}

To bridge the gap between our static analysis and the full dynamic behavior of the system shown in Figure~\ref{fig2}, we extend our steady framework to the oscillatory regime using a quasi-steady approximation. We solve our model at different wall positions over a cycle and average the outflow at each step to obtain the net flow averaged on one cycle \(\langle Q^*\rangle\). Figure~\ref{fig5}(b) compares this quasi-steady model to the numerical simulations. The model accurately captures the relationship between the net flow $\langle Q^*\rangle$ and the visco-elastic number $\eta$ in the high-density limit (Figure~\ref{fig5}(b)). Furthermore, the effect of density remains captured by the scaling $Q \phi/(\phi+k \eta)$ (Figure~\ref{fig5}(b) inset). 
Taken together, this indicates that the effect of $\phi$ and $\eta$ on the flow rate is rooted in quasi-static interactions between the flow and the leaflet deformations. However, we could not find any quasi-static behavior accounting for the effect of the channel length $L_x$ independently of $\phi$, as already noted in Figure~\ref{fig2}(d). As shown in Figure~\ref{fig5}(c), our continuum model fits with the simulations in the limit of long channels but both our continuum model and steady simulations are insensitive to channel length variations (not shown here). This points to a distinct influence of leaflet deformation dynamics.

%Yet, we see in Fig.~\ref{fig5}c that these results do not capture the decrease of the net flow with the decrease of the channel length observed in Fig~\ref{fig2}. This points to a distinct influence of leaflet deformation dynamics, even though the net flow rate remains insensitive to the excitation frequency (Fig.~\ref{fig1}) \sg{I'm not sure I get the sense of this last sentence}.

\begin{figure}[t!]
    \centering
    \includegraphics[width=0.99\textwidth]{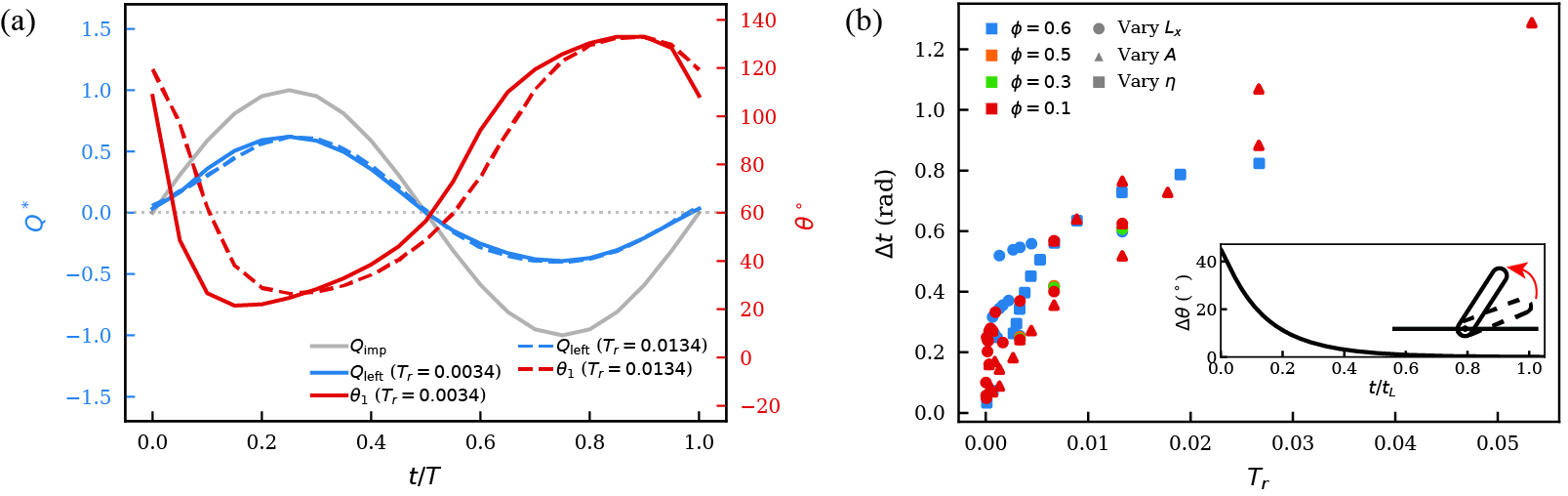}
    \caption{Transient dynamics and the scaling of the leaflet phase lag. \textbf{(a)} Temporal effect of the imposed flow from the upper wall, the instantaneous left channel outflow $Q_{\text{left}}$, and the angular deflection of the left most leaflet $\theta_1$ over a normalized oscillation cycle $t/T$. For two values of the relaxation timescale ($T_r=$0.0034, and 0.0134), where the structural response $\theta_1$ exhibits highly nonlinear dynamics and a distinct phase lag relative to the harmonic hydrodynamic forcing for $T_r=0.0134$. \textbf{(b)} The leaflet phase lag $\Delta t$ with the dimensionless timescale $T_r = (2 D^2 \eta)/(A L_x)$, representing the ratio of the viscous relaxation time of the leaflet to the wall oscillation period. The Phase lag collapses independently of the channel length $L_x = [14.9D:60D]$, oscillation amplitude $A = [0.05D:0.2D]$, elasto-viscous number $\eta = [10^{-4}:0.02]$, and density $\phi = [0.1:0.6]$. This dynamic delay shows the departure from quasi-steady predictions where at large $T_r$, results  a transient configurations that diminish the net rectified transport.  }
    \label{fig6}
\end{figure}

Despite the low Reynolds number value, our time-resolved simulations reveal a notable phase lag between the upper plate motion and the leaflets. Figure~\ref{fig6}(a) highlights the contrast between outflow dynamics and leaflet deformation by comparing the left channel outflow $ Q_{\text{left}}$ with the leftmost leaflet deformation $\theta_1$ during one oscillation cycle $T$. While both the left channel outflow and the leaflet deformation oscillate at the driving frequency of the plate, $\theta_1$ exhibits much more nonlinear dynamics and a distinct phase lag relative to the plate's motion.  
The leaflet response to the flow relies on a viscoelastic time scale
\begin{equation}
  t_{leaflet} = \dfrac{\mu L_p^2}{K}.
\end{equation}
This is the typical relaxation time of a leaflet initially deflected by an angle $\Delta \theta$ from its equilibrium position (see inset in Figure~\ref{fig6}(b)). This relaxation time may induce a phase lag between the leaflet response and the plate oscillation, whenever $t_{leaflet}$ becomes large enough compared to $t_{wall} = 1 / \omega$ and regardless of the value of the Reynolds number.

We quantify the phase lag between the left most leaflet and the imposed flow using a time-domain zero-crossing metric. Specifically, for a given oscillation cycle, we measure the temporal delay $\Delta t = t_2 - t_1$, where $t_1$ is the specific instant the imposed top wall flow crosses zero, and $t_2$ is the subsequent instant the left most leaflet passes through its equilibrium rest angle ($\theta_0$) at the end of the contraction phase. This delay is then normalized by the oscillation period $T$ and plotted in Figure~\ref{fig6}(b) as a function of the time ratio
\begin{equation}
  T_r = \dfrac{t_{leaflet}}{t_{wall}}= \dfrac{2 D^2 \eta}{A L_x}.
\end{equation}
Over large variations of the leaflet density $\phi$, the length of the channel $L_x$, the elasto-viscous number $\eta$ and the amplitude of contraction $A$, the figure confirms that the phase lag scales with $T_r$.

The expression of $T_r$ indicates that the system should depart from a quasi-static behavior when $L_x$ becomes small, consistent with our observations in Figure~\ref{fig5}(c).
It also predicts that time-dependent effects should be stronger for large values of $\eta$, as observed in our simulations.
Interestingly, these dynamical effects are independent of the Reynolds number. They do not explicitly depend on the oscillation frequency either. In contrast, they are entirely controlled by the geometry and elastoviscous parameter.

%To investigate the physical origin of this phase lag, we characterized the relaxation timescale of a leaflet in a viscous fluid (Fig.~\ref{fig6}b). We performed numerical simulations of a single leaflet in a channel, initially deflected by an angle $\Delta \theta$ from its equilibrium position. The study shows that the leaflet relaxation scales with \mb{$T_{r}= $} \oa{$T_{r}=\frac{t_{leaflet}}{t_{wall}}=\frac{\mu L_p^2}{K} \frac{\omega}{2 \pi} = \frac{2 D^2 \eta}{A L_x} $}. \sg{so, this is not exactly shown in fig 9b and I think at this point what we want to say is that the leaflet relaxation time corresponds to $t_{leaflet}$, not $T_r$} 
%We calculated the leaflet phase lag across the system parameters by varying the leaflet density $\phi$, the length of the channel $L_x$, the visco-elasto number $\eta$ and the amplitude of contraction $A$. We show in Fig.~\ref{fig6}b that all the leaflet phase lag follow the same scaling with the characteristic relaxation time of the leaflet $T_r$. This demonstrates that deviations from the quasi-static model stem from dynamic leaflet oscillations. When $T_r$ is large, the leaflets cannot fully reorganize within one oscillation cycle, spending part of each half-cycle in a configuration that partially cancels the rectification generated in the other half, resulting in a reduced net flow rate.
%\vspace{0.5cm}
%\sg{(I propose the following rephrasing starting at line 224)}

%This shows a good scaling and this is natural from the system adimensionalization this is the Strouhal.

\section{Conclusion}\label{Conclusion}

In this work, we investigated how an array of flexible asymmetric leaflets rectifies oscillatory flows in a confined low-Reynolds-number channel. Using fully coupled fluid–structure interaction simulations, we demonstrated that flow-induced leaflet reconfiguration breaks the time-reversal symmetry of the Stokes regime and generates a net directional transport over a complete oscillation cycle.

The transport efficiency is governed primarily by the competition between viscous forcing and elastic restoring forces, quantified by the elastoviscous number $\eta$. We showed that rectification is maximized at an intermediate value of $\eta$. For stiff leaflets, deformations remain too small to generate a significant asymmetry between contraction and relaxation phases, whereas excessively soft leaflets undergo large deformations that reduce the geometric asymmetry responsible for transport. We further demonstrated that collective interactions between leaflets enhance rectification, with the strongest transport obtained in the high-density limit.

To rationalize these observations, we proposed a continuum description of the leaflet bed based on a local torque balance and a global angular-momentum closure. The resulting model accurately predicts the torque distribution and the net flow generated in the dense-leaflet regime. A phenomenological finite-density correction further extends the predictive capability of the model across the full range of leaflet densities investigated here. Together, these results show that the dependence of the transport on elasticity and density is largely controlled by quasi-static fluid–structure interactions.

However, we also showed that departures from quasi-static behavior arise when the leaflet relaxation time becomes comparable to the oscillation period. We showed that these dynamic effects can be characterized by a single dimensionless timescale ratio that controls the phase lag between the flow forcing and the leaflet response, regardless of the Reynolds number.

More broadly, our results highlight how collective fluid–structure interactions can generate symmetry breaking and directional transport in oscillatory viscous flows. The framework introduced here provides a basis for understanding transport mechanisms in biological systems featuring compliant internal structures and suggests new strategies for designing passive microfluidic rectifiers and flow-control devices.

\section*{Acknowledgment} The Centre de Calcul Intensif d’Aix-Marseille is acknowledged for granting access to its high performance computing resources. M.B. and J.P. acknowledge funding from the European Research Council under Grant Agreement No. 101117080.

\section*{Declaration of Interests} The authors report no conflict of interest.

\bibliographystyle{jfm}
\bibliography{jfm}

@article{park2018viscous,
  title={Viscous flow in a soft valve},
  author={Park, Keunhwan and Tixier, Aude and Christensen, AH and Arnbjerg-Nielsen, SF and Zwieniecki, MA and Jensen, KH},
  journal={Journal of Fluid Mechanics},
  volume={836},
  pages={R3},
  year={2018},
  publisher={Cambridge University Press}
}

@article{paludan2024elastohydrodynamic,
  title={Elastohydrodynamic interactions in soft hydraulic knots},
  author={Paludan, Magnus V and Dollet, Benjamin and Marmottant, Philippe and Jensen, Kaare H},
  journal={Journal of Fluid Mechanics},
  volume={984},
  pages={A55},
  year={2024},
  publisher={Cambridge University Press}
}

@article{winn2024operating,
  title={Operating principles of peristaltic pumping through a dense array of valves},
  author={Winn, Aaron and Katifori, Eleni},
  journal={Journal of Fluid Mechanics},
  volume={989},
  pages={A18},
  year={2024},
  publisher={Cambridge University Press}
}

@article{winn2026unidirectional,
  title={Unidirectional flow from continuous broken symmetries},
  author={Winn, Aaron and Parmentier, Justine and Katifori, Eleni and Brandenbourger, Martin},
  journal={arXiv preprint arXiv:2603.27474},
  year={2026}
}

@article{francis2009scaling,
  title={Scaling laws for branching vessels of human cerebral cortex},
  author={Francis, Cassot and Frederic, Lauwers and Sylvie, Lorthois and Prasanna, Puwanarajah and Henri, Duvernoy},
  journal={Microcirculation},
  volume={16},
  number={4},
  pages={331--344},
  year={2009},
  publisher={Taylor \& Francis}
}

@article{park2021fluid,
  title={Fluid-structure interactions enable passive flow control in real and biomimetic plants},
  author={Park, Keunhwan and Tixier, Aude and Paludan, Magnus and {\O}stergaard, Emil and Zwieniecki, Maciej and Jensen, Kaare H},
  journal={Physical Review Fluids},
  volume={6},
  number={12},
  pages={123102},
  year={2021},
  publisher={APS}
}

@article{lu2018biaxial,
  title={Biaxial mechanical behavior of bovine saphenous venous valve leaflets},
  author={Lu, Jiaqi and Huang, Hsiao-Ying Shadow},
  journal={Journal of the Mechanical Behavior of Biomedical Materials},
  volume={77},
  pages={594--599},
  year={2018},
  publisher={Elsevier}
}

@article{levin2024asymmetric,
  title={Asymmetric fluid flow in helical pipes inspired by shark intestines},
  author={Levin, Ido and Sadaba, Naroa and Nelson, Alshakim and Keller, Sarah L},
  journal={Biophysical Journal},
  volume={123},
  number={3},
  pages={21a},
  year={2024},
  publisher={Elsevier}
}

@book{kruger_lattice_2017,
	location = {Cham},
	title = {The Lattice Boltzmann Method: Principles and Practice},
	isbn = {978-3-319-44647-9 978-3-319-44649-3},
	url = {http://link.springer.com/10.1007/978-3-319-44649-3},
	series = {Graduate Texts in Physics},
	shorttitle = {The Lattice Boltzmann Method},
	publisher = {Springer International Publishing},
	author = {Krüger, Timm and Kusumaatmaja, Halim and Kuzmin, Alexandr and Shardt, Orest and Silva, Goncalo and Viggen, Erlend Magnus},
	urlyear = {2024-03-19},
    number={},
	year = {2017},
	langid = {english},
	doi = {10.1007/978-3-319-44649-3},
}

@article{gsell_direct-forcing_2021,
	title = {Direct-forcing immersed-boundary method: A simple correction preventing boundary slip error},
	volume = {435},
	issn = {00219991},
	url = {https://linkinghub.elsevier.com/retrieve/pii/S0021999121001601},
	doi = {10.1016/j.jcp.2021.110265},
	shorttitle = {Direct-forcing immersed-boundary method},
	pages = {110265},
	journal = {Journal of Computational Physics},
	shortjournal = {Journal of Computational Physics},
	author = {Gsell, Simon and Favier, Julien},
	urlyear = {2024-03-19},
	year = {2021},
	langid = {english},
}

@article{gsell_lattice-boltzmann_2021,
	title = {Lattice-Boltzmann simulation of creeping generalized Newtonian flows: Theory and guidelines},
	volume = {429},
	issn = {00219991},
	url = {https://linkinghub.elsevier.com/retrieve/pii/S0021999120307178},
	doi = {10.1016/j.jcp.2020.109943},
	shorttitle = {Lattice-Boltzmann simulation of creeping generalized Newtonian flows},
	pages = {109943},
	journal = {Journal of Computational Physics},
	shortjournal = {Journal of Computational Physics},
	author = {Gsell, Simon and D'Ortona, Umberto and Favier, Julien},
	urlyear = {2024-03-19},
	year = {2021-03},
	langid = {english},
}

@article{moore_lymphatic_2018,
	title = {Lymphatic System Flows},
	volume = {50},
	issn = {0066-4189, 1545-4479},
	url = {https://www.annualreviews.org/doi/10.1146/annurev-fluid-122316-045259},
	doi = {10.1146/annurev-fluid-122316-045259},
	pages = {459--482},
	number = {1},
	journal = {Annual Review of Fluid Mechanics},
	shortjournal = {Annu. Rev. Fluid Mech.},
	author = {Moore, James E. and Bertram, Christopher D.},
	year = {2018},
	langid = {english},
}

@article{brandenbourger_tunable_2020,
	title = {Tunable flow asymmetry and flow rectification with bio-inspired soft leaflets},
	volume = {5},
	issn = {2469-990X},
	url = {https://link.aps.org/doi/10.1103/PhysRevFluids.5.084102},
	doi = {10.1103/PhysRevFluids.5.084102},
	pages = {084102},
	number = {8},
	journal = {Physical Review Fluids},
	shortjournal = {Phys. Rev. Fluids},
	author = {Brandenbourger, M. and Dangremont, A. and Sprik, R. and Coulais, C.},
	urlyear = {2024-03-19},
	year = {2020-08-10},
	langid = {english},
}

@article{gsell_multigrid_2020,
	title = {Multigrid dual-time-stepping lattice Boltzmann method},
	volume = {101},
	issn = {2470-0045, 2470-0053},
	url = {https://link.aps.org/doi/10.1103/PhysRevE.101.023309},
	doi = {10.1103/PhysRevE.101.023309},
	pages = {023309},
	number = {2},
	journal = {Physical Review E},
	shortjournal = {Phys. Rev. E},
	author = {Gsell, Simon and D'Ortona, Umberto and Favier, Julien},
	urlyear = {2024-03-19},
	year = {2020},
	langid = {english},
}

@article{alvarado_nonlinear_2017,
	title = {Nonlinear flow response of soft hair beds},
	volume = {13},
	issn = {1745-2473, 1745-2481},
	url = {https://www.nature.com/articles/nphys4225},
	doi = {10.1038/nphys4225},
	pages = {1014--1019},
	number = {10},
	journal = {Nature Physics},
	shortjournal = {Nature Phys},
	author = {Alvarado, José and Comtet, Jean and De Langre, Emmanuel and Hosoi, A. E.},
	urlyear = {2024-03-19},
	year = {2017},
	langid = {english},
}

@article{wehner2016integrated,
  title={An integrated design and fabrication strategy for entirely soft, autonomous robots},
  author={Wehner, Michael and Truby, Ryan L and Fitzgerald, Daniel J and Mosadegh, Bobak and Whitesides, George M and Lewis, Jennifer A and Wood, Robert J},
  journal={nature},
  volume={536},
  number={7617},
  pages={451--455},
  year={2016},
  publisher={Nature Publishing Group UK London}
}

@article{garcia2025spontaneous,
  title={Spontaneous emergence of solitary waves in active flow networks},
  author={Garc{\'\i}a, Rodrigo Fern{\'a}ndez-Quevedo and Antunes, Gon{\c{c}}alo Cruz and Harting, Jens and Stark, Holger and Valeriani, Chantal and Brandenbourger, Martin and Mazo, Juan Jos{\'e} and Malgaretti, Paolo and Ruiz-Garc{\'\i}a, Miguel},
  journal={arXiv preprint arXiv:2511.13448},
  year={2025}
}

@article{martinez2024fluidic,
  title={The fluidic memristor as a collective phenomenon in elastohydrodynamic networks},
  author={Mart{\'\i}nez-Calvo, Alejandro and Biviano, Matthew D and Christensen, Anneline H and Katifori, Eleni and Jensen, Kaare H and Ruiz-Garc{\'\i}a, Miguel},
  journal={Nature Communications},
  volume={15},
  number={1},
  pages={3121},
  year={2024},
  publisher={Nature Publishing Group UK London}
}

@article{mosadegh2010integrated,
  title={Integrated elastomeric components for autonomous regulation of sequential and oscillatory flow switching in microfluidic devices},
  author={Mosadegh, Bobak and Kuo, Chuan-Hsien and Tung, Yi-Chung and Torisawa, Yu-suke and Bersano-Begey, Tommaso and Tavana, Hossein and Takayama, Shuichi},
  journal={Nature physics},
  volume={6},
  number={6},
  pages={433--437},
  year={2010},
  publisher={Nature Publishing Group UK London}
}

@book{10.1039/9781782628491,
    author = {Duprat, Camille and Stone, Howard},
    title = {Fluid–Structure Interactions in Low-Reynolds-Number Flows},
    publisher = {The Royal Society of Chemistry},
    year = {2015},
    month = {11},
    isbn = {978-1-84973-813-2},
    doi = {10.1039/9781782628491},
    url = {https://doi.org/10.1039/9781782628491},
}

@misc{jambon-puillet_dense_2025,
	title = {Dense array of elastic hairs obstructing a fluidic channel},
	url = {http://arxiv.org/abs/2501.01875},
	doi = {10.48550/arXiv.2501.01875},
	number = {{arXiv}:2501.01875},
	publisher = {{arXiv}},
	author = {Jambon-Puillet, Etienne},
	urlyear = {2025-01-07},
	year = {2025},
	eprinttype = {arxiv},
	eprint = {2501.01875 [cond-mat]},
}

@article{guo_discrete_2002,
	title = {Discrete lattice effects on the forcing term in the lattice Boltzmann method},
	volume = {65},
	rights = {http://link.aps.org/licenses/aps-default-license},
	issn = {1063-651X, 1095-3787},
	url = {https://link.aps.org/doi/10.1103/PhysRevE.65.046308},
	doi = {10.1103/PhysRevE.65.046308},
	pages = {046308},
	number = {4},
	journal = {Physical Review E},
	shortjournal = {Phys. Rev. E},
	author = {Guo, Zhaoli and Zheng, Chuguang and Shi, Baochang},
	urlyear = {2025-04-03},
	year = {2002},
	langid = {english},
}

@article{thomazo_collective_2020,
	title = {Collective stiffening of soft hair assemblies},
	volume = {102},
	issn = {2470-0045, 2470-0053},
	url = {http://arxiv.org/abs/2002.02834},
	doi = {10.1103/PhysRevE.102.010602},
	pages = {010602},
	number = {1},
	journal = {Physical Review E},
	shortjournal = {Phys. Rev. E},
	author = {Thomazo, Jean-Baptiste and Lauga, Eric and Révérend, Benjamin Le and Wandersman, Elie and Prevost, Alexis Michel},
	urlyear = {2025-08-12},
	year = {2020-07-02},
	eprinttype = {arxiv},
	eprint = {2002.02834 [cond-mat]},
}

\end{document}